# Nonlinear Polarization and Efficiency *Droop* in Hexagonal InGaN/GaN Disk-in-Wire LEDs

Vinay Uday Chimalgi, Md Rezaul Karim Nishat, and Shaikh Shahid Ahmed, *Senior Member*, IEEE

*Abstract*—Recent studies suggest that piezoelectric polarization can play an important role in determining the electronic and optical properties of nanoscale nitride heterostructures. Among a few models available, recent first-principles calculations performed by Prodhomme *et al.* provide a simple yet accurate description of linear and nonlinear piezoelectric coefficients in reduced dimensionality structures having wurtzite crystal symmetry. In this paper, first, within a *fully atomistic* VFF-$sp^3s^*$ tight-binding framework, we employ the model proposed by Prodhomme *et al.* to evaluate the importance of *nonlinear piezoelectricity* on the single-particle electronic states and interband optical transitions in a recently reported *hexagon* shaped $In_{0.25}Ga_{0.75}N$/GaN disk-in-wire LED. The microscopically determined transition parameters are then incorporated into a TCAD toolkit to investigate how atomicity and the net polarization field affect the internal quantum efficiency of the LED and lead to a degraded *efficiency droop* characteristic.

*Index Terms*—nonlinear piezoelectricity, tight-binding, optical anisotropy, disk-in-wire LED, efficiency *droop*.

## I. INTRODUCTION

NANOENGINEERED InGaN heterostructures are promising candidates for efficient, full-spectrum optical emitters for use in both conventional (e.g. lasers, solid-state lighting) and novel (e.g. near-field photolithography, single phonon based quantum cryptography, diagnostic medicine and biological imaging) application domains [1]. Traditionally, most InGaN heterostructures have been demonstrated in two forms: a) quasi 2-D quantum wells (QW) grown using molecular beam epitaxy; and b) quasi 0-D self-assembled quantum dots (QD) [2][3]. QDs, as compared to QWs, benefit from the presence of a relaxed (less-strained) active layer that leads to a reduced quantum confined Stark effect (QCSE). Nevertheless, recent studies show that the electronic and optical properties of these quasi zero-dimensional systems can be adversely affected by an intricate interplay of the quantum confinement effects and the *atomicity* in the underlying crystal lattice and interfacial discontinuities [4][5].

Very recently, expitaxially grown InGaN heterostructures realized in *disk-in-wire* architectures have been hailed as a breakthrough technology for applications in quantum photonics [6]. Various experimental studies demonstrate that these novel structures: a) are highly stable possessing greater reproducibility; b) can be grown free of extended defects, which allows the use of higher indium content and a wider spectral tuning; and c) due to the presence of greater geometrical symmetry, are subject to a lesser degree of internal electro-mechanical fields. In wurtzite heterostructures, compared to zincblende systems, the effects of polarization fields is strong, where, in addition to the off-diagonal strain tensors, biaxial strain induces a piezoelectric field parallel to the *C* axis ([0001] direction). Additionally, recent studies indicate that *nonlinear* ($2^{nd}$-order) piezoelectric polarization can also influence the electronic and optical properties of nanoscale III-V heterostructures [5][7][8]. For strained InAs/GaAs QDs, it was shown that, for large epitaxial strains, the $1^{st}$-order and the $2^{nd}$-order contributions are of comparable magnitude [9]! In this paper, considering a recently reported *hexagon* shaped InGaN/GaN disk-in-wire LED [6] as a reference device, we address the following three questions: a) Can the InGaN active disk buried in a GaN wire be *completely* strain-free?; b) Should one, while designing these devices, take into account the contribution of nonlinear polarization?; and c) How do various internal fields affect the internal quantum efficiency and associated *droop* (roll-over) characteristics in InGaN LEDs?

## II. THEORY AND MODELS USED

Nanostructured InGaN optical emitters, in addition to quantum confinement, are subject to: a) symmetry lowering due to fundamental crystal atomicity and material discontinuity at the interfaces; b) non-homogeneous strain distribution; c) strain-induced piezoelectric polarization, $P_{PZ}$; and d) spontaneous polarization, $P_{SP}$. For an accurate treatment of these internal fields, in this work, the numerical analysis has been carried out using a combination of valence force-field molecular mechanics (VFF-MM), 10-band $sp^3s^*$ atomistic tight-binding (TB), and a TCAD based quantum-corrected drift-diffusion transport framework as available in our in-house QuADS 3-D (quantum atomistic device simulation in three dimensions) simulator [10]. Note that, tight-binding is a local basis representation, which naturally deals with finite device sizes, alloy-disorder and hetero-interfaces and results in computationally tractable sparse matrices. The overall simulation strategy has been described elsewhere [5][11]. After creating the entire structure from a basis set, the atom positions are relaxed and the resulting strain fields are calculated via the valence force-field (VFF) method using the

Manuscript prepared August 20, 2014. This work was supported by National Science Foundation Grant No. 1102192.

The authors are with the Department of Electrical and Computer Engineering, Southern Illinois University at Carbondale, IL 62901 USA (phone: 618-453-7630; fax: 618-453-7972; email: ahmed@siu.edu).

TABLE I. POLARIZATION CONSTANTS (IN C/M²) FOR GaN AND InN

| | $P_{PZ}^{bulk}$ (expt.)[13] | | | $P_{PZ}$ (ab initio 1st-order)[8] | | | $P_{PZ}$ (ab initio 2nd-order)[8] | | | | $P_{SP}$[8] |
|---|---|---|---|---|---|---|---|---|---|---|---|
| | $e_{15}$ | $e_{31}$ | $e_{33}$ | $e_{15}$ | $e_{31}$ | $e_{33}$ | $e_{311}$ | $e_{312}$ | $e_{313}$ | $e_{333}$ | |
| GaN | 0.326 | -0.527 | 0.895 | -0.31 | -0.44 | 0.75 | 6.2 | 3.3 | 0.4 | -21.4 | -0.027 |
| InN | 0.264 | -0.484 | 1.06 | -0.43 | -0.59 | 1.14 | 4.8 | 3.7 | 0.5 | -18.6 | -0.035 |

Keating potentials. The strain parameters used in this work are taken from Ref. [12] and validated through the calculation of Poisson ratio of the underlying materials. The small thermal strain contribution is neglected. The spontaneous polarization is strain-independent and arises from fundamental asymmetry of the crystal structure. In contrast, the piezoelectric polarization is obtained from the diagonal and shear components of the anisotropic *atomistic* strain fields [13]. To account for the 2nd-order piezoelectric field, a recent model of Prodhomme *et al.* [8] has been implemented in this work. In Ref. [8], to obtain the second-order piezoelectric coefficients, the authors have used a finite difference technique in conjunction with density functional perturbation theory (DFPT) within the local density approximation (LDA). The calculations were carried out with the *Abinit* toolkit using the Troullier-Martins (TM) pseudopotentials. Here, the net piezoelectric polarization (including the 1st-order and the 2nd-order contributions) along the [0001] direction in a wurtzite crystal is given by:

$$P_{PZ,z}^{net} = e_{31}\left(\varepsilon_{xx} + \varepsilon_{yy}\right) + e_{33}\varepsilon_{zz} + \frac{1}{4}\left(e_{311} + e_{312}\right) \times \left(\varepsilon_{xx} + \varepsilon_{yy}\right)^2 + \frac{1}{2}e_{333}\varepsilon_{zz}^2 + 2e_{313}\varepsilon_{xx}\varepsilon_{zz}.$$

The polarization constants (in C/m²) used in this study are listed in Table I. Note that: a) the second-order polarization coefficients are almost an order of magnitude higher than the linear counterparts; and b) of all the coefficients, $e_{333}$ has the largest magnitude and can induce significant second-order effects especially in materials with a rather small linear coefficient $e_{33}$ and large epitaxial strain. Also, note that, compared to an earlier 2nd-order model [7] proposed by Pal *et al.* (that was implemented in a recent work [5] of the current authors), the new model proposed by Prodhomme *et al.* reported (from symmetry considerations) a *zero* value for the coefficient $e_{133}$ [14]. Next, the polarization induced potential is obtained by solving the 3-D Poisson equation on an atomistic grid using the open source PETSc toolkit [15]. Next, the single-particle eigenvalues, wave functions, and the interband optical transition rates are calculated using an empirical nearest-neighbor $sp^3s^*$ tight-binding model as implemented in the open source NEMO 3-D simulator [16]. Finally, the microscopically determined optical transition parameters are incorporated in the Synopsys' *Sentaurus* toolkit to obtain the terminal characteristics of the LED and investigate how various internal fields affect the internal quantum efficiency of the device.

## III. RESULTS AND DISCUSSION

Fig. 1 shows the schematic of the simulated InGaN *disk-in-wire* LED. The core atomistic simulation domain assumes a *hexagonal* geometry for the wurtzite crystal system. The GaN nanowire is oriented in the [0001] direction (*c*-axis) with longest diagonal, $d$~14 nm (side, $s$~7 nm) and height, $h$~100 nm. The In$_{0.25}$Ga$_{0.75}$N quantum disk is positioned at the center of the GaN nanowire and has a height of $h_d$~3 nm. Looking at the strain profiles, as shown in Fig. 2, both the hydrostatic $\left(\varepsilon_{xx} + \varepsilon_{yy} + \varepsilon_{zz}\right)$ and the biaxial $\left(\varepsilon_{xx} + \varepsilon_{yy} - 2\varepsilon_{zz}\right)$ components were found to be compressive within the disk and tensile in the surrounding material matrix.

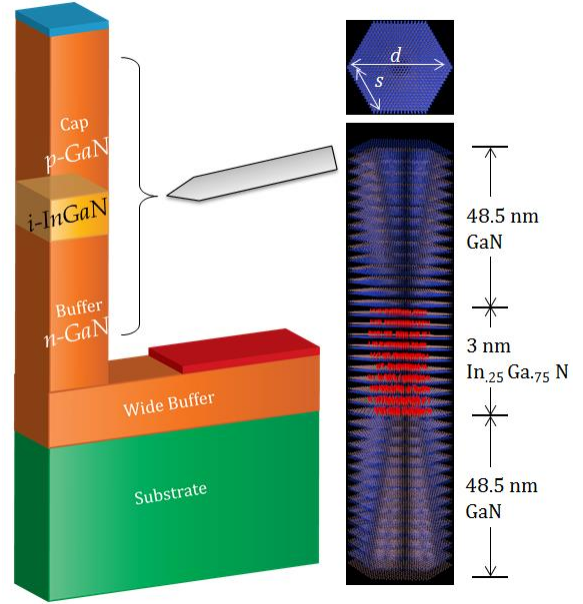

Fig. 1. (left) Schematic representation of the simulated InGaN quantum disk-in-wire LED. (right) Atomicity of the active region.

Fig. 3(a) shows the polarization-induced piezoelectric potential distribution (for different model configurations) along the growth direction and through the center of the disk. The salient features one can extract from this Figure are as follows: a) The 2nd-order piezoelectric contribution is significantly smaller than the 1st-order counterpart, magnitude being only ~18.3 mV at the InGaN/GaN interfaces and almost negligible in the GaN substrate and cap layers; and b) The net potential peaks at the interfaces at ~134 mV, tends to be symmetric around the center of the dot (along the growth direction), and, as seen in Fig. 3(b), features strong dipole formation in the *x-z* plane.

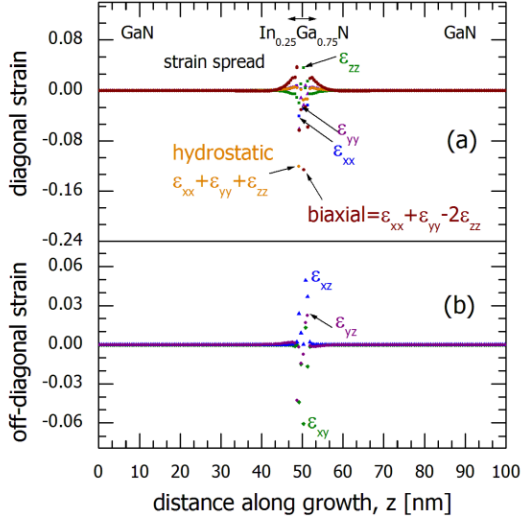

Fig. 2. (a) Atomistic diagonal, biaxial, and hydrostatic strain profiles along the growth ([0001]) direction through the center of the structure. Strain is seen to penetrate deep (~15 nm) into the GaN matrix. (b) Off-diagonal strain variations along the [0001] direction.

Fig. 4 shows the HOMO ($E_V$) and the LUMO ($E_C$) wavefunctions projected on the *x-y* plane in the InN quantum disk. Here, it is clear that, due to the atomistic and random nature of the distribution of atoms, even in the absence of internal fields (depicted as WO), isotropy is never achieved. Inclusion of atomistic strain and electrostatic fields in the calculation leads to a *strong* localization in the HOMO wavefunctions, and, therefore, lowers the overall electronic symmetry of the system. The bottom row in this Figure shows the single-particle bandgap, $E_G$, of the LED structure, which, in general, agrees well with the experimentally measured value (~2.73 eV at 300°K) as reported in Ref. [6].

The influence of the internal fields on the single-particle bandgap is summarized in Fig. 5. Looking at this Figure, we can infer that: a) both quantum confinement and strain lead to a large blue-shift in the bandgap, strain being the stronger contributor of the two; b) pyroelectric contribution is negligible; c) the 1st-order piezoelectricity *alone* introduces a significant redshift and cannot be ignored; d) the bulk or experimental (*expt*) piezoelectric constant overestimates the degree of red-shift, as compared to the same obtained from first-principles calculations for nanostructures; and e) 2nd-order piezoelectricity, while opposing the 1st-order counterpart, is very weak in magnitude and may be ignored in the calculation of the single-particle energy bandgap.

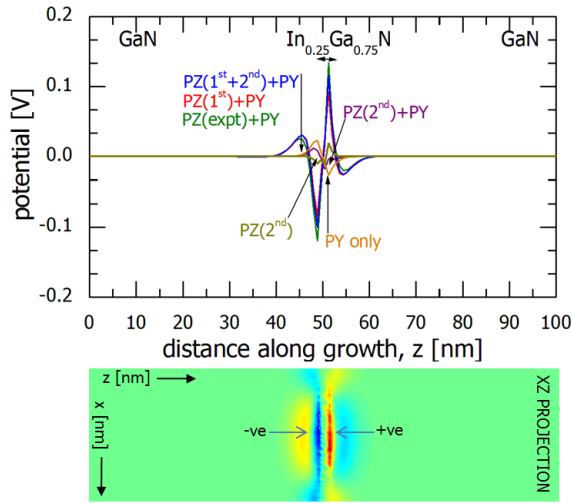

Fig. 3. (a) Polarization-induced piezoelectric potential distribution along the growth direction and through the center of the disk. (b) Potential distribution projected on the *x-z* plane showing formation of the dipole.

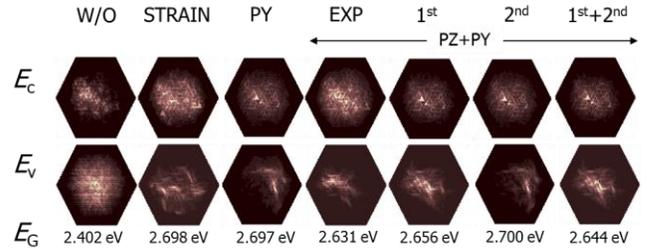

Fig. 4. HOMO and LUMO wavefunctions featuring strong localization and anisotropy due to quantum atomicity, strain, and the piezoelectric fields.

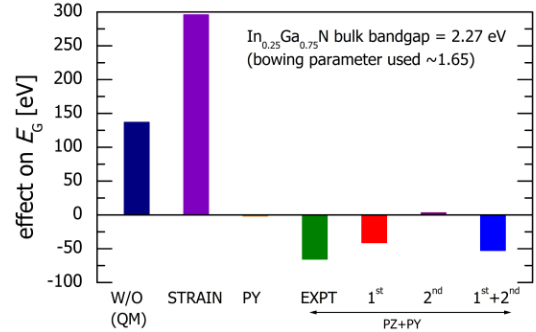

Fig. 5. Energy bandgap as a function of atomicity (unrelaxed lattice), strain, and various polarization models (in increased order of accuracy).

Fig. 6 shows the polar plots of the interband optical transition rates between ground hole (HOMO) and ground electron (LUMO) states. It is found that, all internal fields (as compared to the WO case) break the isotropy in the emission characteristic with varying degrees. Overall, the *in-plane polarization anisotropy* was found to be ~0.91, due mainly to strong localization and spatial irregularity in the wavefunctions. Note that, the 2$^{nd}$-order piezoelectricity, while found to be insignificant in the calculation of the energy bandgap, renders a non-vanishing effect in retaining the symmetry of the system (*pushes* the spectrum clockwise).

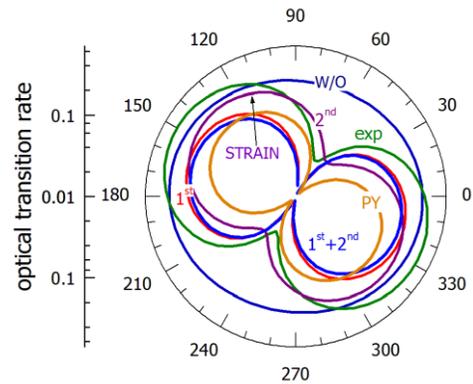

Fig. 6. Polar plots of interband optical transition rates between the HOMO and the LUMO states in the InGaN quantum disk.

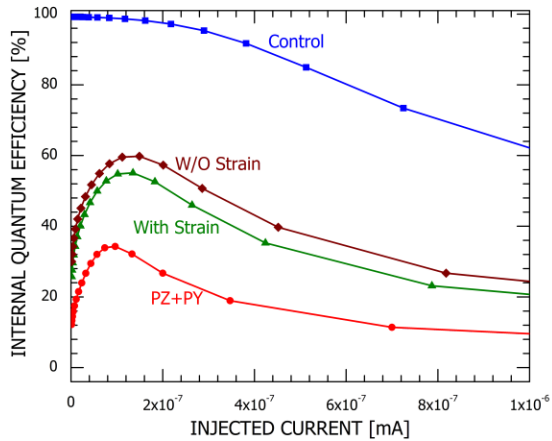

Fig. 7. Simulated internal quantum efficiency (IQE) as a function of injected current for the InGaN disk-in-wire LED. Here, *Control* refers to a device without any structural disorder and $k_{sp}$ =1.

Next, Synopsys' Sentaurus TCAD toolkit is used to simulate the optical characteristics of the disk-in-wire LED. Here, the core active region comprises of an *undoped* 3 nm thick In$_{0.25}$Ga$_{0.75}$N disk sandwiched between an *n*-doped GaN buffer region and a *p*-doped GaN cap layer, both having a base length, $b$~14 nm and height, $h$~48.5 nm. Note that, the effect of internal fields, which is characterized by a strong suppression of interband optical transition rate, is modeled via a *scaling factor* ($k_{sp}$) for the optical matrix element accessible in the TCAD toolkit. The $k_{sp}$ values used here are the average transition rates obtained from Fig. 6. Fig. 7 shows the variation of internal quantum efficiency (IQE) as a function of injected current for various models used in the calculation. Here, the rollover in efficiency (*droop*), even in the absence of any structural disorder (defects, SRH centers, strain and polarization fields, depicted as *Control* structure), is due mainly to carrier loss from within the confined disk region. Importantly, it is found that, inclusion of the internal polarization fields in the calculation: a) reduces the IQE drastically; and b) further degrades the *droop* characteristic by inducing the same at a smaller injected current value.

## IV. Conclusion

InGaN/GaN disk-in-wire heterostructures, although realized with a greater geometrical symmetry and a more relaxed lattice compared to the self-assembled QD counterparts, are still subject to non-negligible epitaxial strain and compositional and interfacial atomicity. For a recently reported device, full atomistic calculations of the electronic structure reveal a strong dependence of the bandgap on the internal fields and feature large growth-plane optical polarization anisotropy. The 2$^{nd}$-order piezoelectricity, while minimally affecting the energy bandgap, renders a non-vanishing effect in retaining the symmetry of the system. Overall, the net polarization field was found to degrade the efficiency *droop* characteristic and, therefore, must be taken into account in modeling these novel LED structures.

## Acknowledgment

This work was supported by the U.S. National Science Foundation Grant No. 1102192. Access to the Synopsys TCAD toolset via SIU site is also acknowledged.

## References

[1] M. Razeghi, "III-Nitride Optoelectronic Devices: From Ultraviolet Toward Terahertz," *IEEE Photonics Journal*, vol. 3, no. 2, pp. 263–267, April 2011.
[2] C. Adelmann, J. Simon, G. Feuillet, N. T. Pelekanos, B. Daudin and G. Fishman, "Self-assembled InGaN quantum dots grown by molecular-beam epitaxy," *Appl. Phys. Lett.*, vol. 76, pp.1570–1572, 2000.
[3] J. Kalden, C. Tessarek, K. Sebald, S. Figge, C. Kruse, D. Hommel, and J. Gutowski, "Electroluminescence from a single InGaN quantum dot in the green spectral region up to 150 K," *Nanotechnology*, vol. 21, 015204, 2010.
[4] S. Ahmed, S. Islam, and S. Mohammed, "Electronic Structure of InN/GaN Quantum Dots: Multimillion Atom Tight-Binding Simulations," *IEEE Trans. Electron Devices*, vol. 57, no. 1, pp. 164–173, 2010.
[5] K. Yalavarthi, V. Chimalgi and S. Ahmed, "How Important is Nonlinear Piezoelectricity in Wurtzite GaN/InN/GaN Disk-in-Nanowire LED Structures?" *Opt. Quant. Electron.*, vol. 46, pp. 925–933, 2014.
[6] S. Deshpande, J. Heo, A. Das, and P. Bhattacharya, "Electrically driven polarized single-photon emission from an InGaN quantum dot in a GaN nanowire," *Nature Communications*, vol. 4, Article number. 1675, DOI: 10.1038/ncomms2691.
[7] J. Pal, G. Tse, V. Haxha, and M. A. Migliorato, "Second-order piezoelectricity in wurtzite III-N semiconductors," *Phys. Rev. B*, vol. 84, 085211, 2011.
[8] P-Y Prodhomme, A. Beya-Wakata, and G. Bester, "Nonlinear piezoelectricity in wurtzite semiconductors," *Phys. Rev. B*, vol. 88, 121304(R), 2013.
[9] G. Bester, A. Zunger, X. Wu, and D. Vanderbilt, "Effects of linear and nonlinear piezoelectricity on the electronic properties of InAs/GaAs quantum dots", *Phys. Rev. B*, vol. 74, 081305, 2006.
[10] S. Ahmed, K. Yalavarthi, V. Gaddipati, A. Muntahi, S. Sundaresan, S. Mohammed, S. Islam, R. Hindupur, D. John, and J. Ogden, "Quantum Atomistic Simulations of Nanoelectronic Devices using QuADS," In *Nano-Electronic Devices: Semiclassical and Quantum Transport Modeling*, Springer, Edited by D. Vasileska and S. M. Goodnick, pp. 405–441, 2011.
[11] S. Sundaresan, V. Gaddipati, and S. Ahmed, "Effects of Spontaneous and Piezoelectric Polarization Fields on the Electronic and Optical Properties in GaN/AlN Quantum Dots: Multimillion-Atom $sp^3d^5s^*$ Tight-Binding Simulations," *Int. Journal of Numerical Modeling*, Wiley, DOI: 10.1002/jnm.2008, published online 2 Jun 2014.
[12] T. Saito, Y. Arakawa, "Electronic structure of piezoelectric In$_{0.2}$Ga$_{0.8}$N quantum dots in GaN calculated using a tight-binding method", *Physica E: Low-Dimensional Systems and Nanostructures*, vol. 15, pp. 169–181, 2002.
[13] M. Winkelnkemper, A. Schliwa, and D. Bimberg, "Interrelation of structural and electronic properties in In$_x$Ga$_{1-x}$N/GaN quantum dots using an eight-band k•p model," *Phys. Rev. B*, vol. 74, 155322, 2006.
[14] Personal communication with Professor Gabriel Bester
[15] PETSc (*P*ortable, *E*xtensible *T*oolkit for *S*cientific *C*omputation, pronounced *PET-see*);
Freely available at: http://www.mcs.anl.gov/petsc/.
[16] G. Klimeck, S. Ahmed, N. Kharche, H. Bae, S. Clark, B. Haley, S. Lee, M. Naumov, H. Ryu, F. Saied, M. Prada, M. Korkusinski, and T. Boykin, "Part I: Atomistic Simulation of Realistically Sized Nanodevices Using NEMO 3-D", *IEEE Trans. Electron Devices*, vol. 54, 9, pp. 2079–89, 2007.